# If Global or Local Investor Sentiments are Prone to Developing an Impact on Stock Returns, is there an Industry Effect?

**Jing Shi and Marcel Ausloos (*)**

**School of Business, University of Leicester, Brookfield Campus, Leicester, LE2 1RQ, United Kingdom**

**Tingting Zhu**
**Faculty of Business and Law, Hugh Aston Building, De Montfort University, Leicester, LE1 9BH, United Kingdom**

**(*) corresponding author: marcel.ausloos@uliege.be**
**ORCID: 0000-0001-9973-0019**

We confirm that this article contains a Data Availability Statement (see section 4 ). We confirm that we include how available data was and can be obtained ; see references section also.

Abstract

This paper investigates the heterogeneous impacts of either Global or Local Investor Sentiments on stock returns. We study 10 industry sectors through the lens of 6 (so called) emerging countries: China, Brazil, India, Mexico, Indonesia and Turkey, over the 2000 to 2014 period. Using a panel data framework, our study sheds light on a significant effect of **Local Investor Sentiments** on expected returns for basic materials, consumer goods, industrial, and financial industries. Moreover, our results suggest that from **Global Investor Sentiments** alone, one cannot predict expected stock returns in these markets.

## 1. Introduction

To consider investor sentiments (IS), as a persistent vehicle to capture market fluctuations, is a modern research field (Neal and Wheatley, 1998), but had been already considered by Keynes (1936). Fisher and Statman (2000, 2003) even distinguished « Wall Street strategists from individual investors or newsletter writers ». That was followed by Brown and Cliff (2005), Chang et al. (2011), and Baker et al. (2012). Schmeling (2009) studied IS and stock returns (SR) in 18 industrialized countries and Baker and Wurgler (2006) found that IS have « larger effects on securities whose valuations are highly subjective and difficult to arbitrage ». Since equities from different industries have different characteristics, the impacts of investor sentiments should be expected to vary among industries, as promoted by Chen, Chen, and Lee (2013). In fact, empirical and theoretical evidences (Bu and Li, 2014; Dhesi and Ausloos, 2016) recently show that investors' (irrational or not) emotions, tied to economic or other news, seem to discrepantly sway the future returns of stocks.

Yet to our knowledge there is only a small number of studies which focus on the correlation between specific industries' expected returns and investor sentiments. Beside the remarkable paper by Chen et al. (2013), only a few scholars have regarded emerging markets as the investigated objects for an IS-SR nexus research.

Our study focusses on industry type correlations with both global or local IS in a few emerging markets. Both local investor sentiments (LIS) and global investor sentiments (GIS) are collected within a panel data set which contains the 2000 to 2014 period for 6 (so called) emerging countries, and 10 industries: basic materials, consumer goods, consumer services, financials, health care, industrials, oil and gas, technology, telecommunications, and utilities. An industry refers to a group of companies share the similar business activities, i.e., producing products, services or close substitutes. The classification schemes reflect different sources, revenues, and the stages of industry cycle, therefore respond differently to information. For each industry, the stock returns are computed from the data of FTSE global price indexes. We use each country's (thus local) consumer confidence index to reflect LIS; we measure the GIS using the State Street Investor Confidence Index (http://www.statestreet.com/ideas/investor-confidence-index.html). We include business cycle as a control variable in our regression framework returns, following McLean and Zhao's (2014) recommendation.

We find that the LIS has a heterogeneous effect on expected stock returns in different industries. In particular, the empirical results suggest that the expected returns of basic materials, consumer goods, industrials and financials industries have significant and positive correlations with LIS. This finding is consistent with most of previous conclusions. However, we find that GIS do not allow to predict expected stock returns. Moreover, our conclusive results also suggest that it is inadequate to only use LIS and GIS to perform a rational anticipation of future stock returns.

Consequently, this study contributes to the literature in three folds. First, the study focuses on the role of industry in the correlation between IS and expected SR. One of our main motivations is to find which industry is the most vulnerable to investors' emotion fluctuations and subsequently which industry is most stable during the bullish or bearish sentiment periods. Second, we take both GIS and LIS into consideration. Third, this paper provides studies through the lens of emerging countries: China, Brazil, India, Mexico, Indonesia and Turkey. We use the data between January 2000 and December 2014 to guarantee the volatility and practicability of the research. To the best of our knowledge, this is the first study to investigate the correlation between investor sentiments and different industries' expected stock returns with the panel data of emerging countries, on an equal footing.

The remainder of the paper is organized as follows. We propose our hypotheses in Section 2. Section 3 outlines the methodology with our comments on relevant variables. Section 4 presents data adjustments. In Section 5, we test our hypotheses and reports the findings. Section 6 serves as a conclusion. There are several Appendices containing « numerical and technical details ».

## 2. Hypotheses development

There are numerous studies which focus on the correlation between stock returns and investor sentiments; details of those previous studies are displayed in Table 1. However, since the existing studies of investor sentiments and stock returns are too numerous to be all mentioned, we only list articles which are representative and frequently cited in the literature. According to Table 1, we can conclude that investor sentiments are proved to be influential to expected stock returns in many countries. We therefore formulate our first hypothesis as follows:

**Hypothesis 1**. LIS has a positive impact on industry stock returns in emerging countries.

Next, let it be observed that there is a scarcity of evidence on the impact of global sentiments. As noted earlier, it is important to investigate global sentiments along with local sentiments. Chang et al. (2011:36) confirmed the impact of GIS across international markets, with the claim that the LIS effect is "just an empirical manifestation of the global effect''.

Therefore, our second hypothesis reads as follows:

**Hypothesis 2**. GIS has a positive impact on industry stock returns in emerging countries, with the effect that more accessible markets will be associated with a stronger GIS.

Notice that Baker et al. (2012) found that both global and local market sentiments have a predictive power on industry returns, global sentiment playing a more significant role. Since only a few researchers perform empirical studies on the relationship(s) between sentiments and stock returns, distinguishing different industries, the research question (RQ) tied to H1 and H2 immediately follows:

**RQ:** If GIS or LIS are prone to developing an impact on SR, is there an industry effect ? – at least in emerging countries and for a considered time interval.

## 3. Methodology and Variables

We employ a classical linear panel fixed-effect regression model (Simpson & Ramchander, 2002; Brown & Cliff, 2006; Baker & Wurgler, 2006; Tsuji, 2006; Schmeling, 2009; Chen et al., 2013) which will be separately conducted for the different industries in order to make some comparison:

$$SR_{it+1} = \beta_{0i} + \beta_1 LIS_{it} + \beta_2 GIS_t + \beta_3 BC_{it} + \varepsilon_{it+1}$$

where $i=1, 2, 3, \ldots, N$ ; $t=1, \ldots, T$; $i$ and $t$ refer to the country number and time, respectively; $N$ is the maximum number of countries so examined. Notice that the error should be considered as taken at the date of the forecast, $t+1$, not the day before.

Let for China, Brazil, India, Mexico, Indonesia and Turkey, the country number to be 1, 2, 3, 4, 5, and 6, respectively. *SR* represents the expected stock returns for a specific industry, $SR_{it+1}$ denotes the industry stock returns for country $i$ at time $t+1$; the explanatory variables include each country's consumer confidence index as a proxy for LIS (*LIS*), the state street investor confidence index as a proxy for the global investor sentiments (*GIS*), and the unemployment rate as a proxy for business cycle (*BC*). Accordingly, $LIS_{it}$, and $BC_{it}$ represent local market sentiments, global sentiments and business cycle factor for country $i$ at time $t$, $GIS_t$ denotes the GIS at time $t$ , while $\beta_1, \beta_2$ $\beta_3$ are the regression coefficients of local sentiments, global sentiments and business cycle respectively. They also represent the marginal effects of those explanatory variables on industry stock returns; $\varepsilon_{i\ t+1}$ is the disturbance term for country $i$ at time $t+1$. An independent and identically distributed (I.I.D.) assumption on $\varepsilon_{i\ t+1}$ indicates that the data is expected to be independent, identically distributed. This assumption allows one to model the joint probability of the data as the product of the probability of individual data points. The parameter $\beta_{0i}$ is a time independent term which allows the possibility of country-specific fixed effects.

## 3.1 Investor Sentiments

LIS and GIS are the dependent variables throughout the empirical tests. Many scholars, e.g. Charoenrook (2003), Lemmon and Portniaguina (2006), and Schmeling (2009), have employed the consumer confidence index (CCI) to perform sentiment studies (Trading economics, 2015). All countries considered in our research have their own official consumer confidence indicator and we have employed each country's official consumer confidence index as a sentiment proxy for local sentiments throughout the paper.

The State Street Investor Confidence Index quantitatively measures the (global) investor sentiments wherever the world country is. This index is built by estimating the changes in institutional investors' position of risky assets and is based on a specialized research model (State Street, 2015). This index relies on the actual financial trading behaviors of institutional investors and demonstrates traders' real feelings about the expected returns.

Figure 1 depicts the Global and Local Investor Sentiments for the time and countries of interest. Notice visually that the global investor sentiment index appears to have a weak correlation with the local investor sentiments indexes: the LIS and GIS indicators appear to move discrepantly during the investigated time interval. For example, the GIS index trends decreased in the 2000-2004 period, while almost all LIS index trends increased during that time.

### 3.2 Industry Stock Returns

The dependent variable is the stock returns of different industry sectors. As company's demand and supply within a given industry are affected by environmental forces and economic performances in much the same way (Bain, 1959; Porter, 1980), their stock returns are correlated. We follow the industry classification benchmark of the Financial Times Security Exchange (henceforth FTSE) group, which divides all traded equities into ten different industry categories: basic materials (*MATS*), consumer goods (*GDS*), consumer services (*SVS*), financials (*FIN*), health care (*HEA*), industrials (*INDU*), oil and gas (*OIL*), technology (*TECH*), telecommunications (*TELE*), and utilities (*UTIL*). In order to measure each industry's SR, we employ the FTSE « Global Monthly Price Index » (GPI) for each industry sector in the six countries and transform GPIs into log returns. Additionally, the currency unit of the collected price indexes is USD, in order to eliminate the influence of exchange rates. It is fair mentioning that there are some limitations to the FTSE Global Index. Firstly, it only employs some specific portfolios to represent the whole market. Secondly, the FSTE price index is not available for several countries' specific industry sectors; in our case, for example, there is no FSTE price index for Mexico technology industry.

### 3.3 The Business Cycles

We control the business cycle for macro-economic impacts on stock returns, using the unemployment rate in each country to capture each country's long-term macro-economic situation. It can be argued that the unemployment rate could only partly reflect a country's long-term macro-economic situation, because it is merely one particular indicator of a country's particular economic condition. Aware of such arguments and limitations, nevertheless, we use the unemployment rate as indicator for measuring the macro-economic status for the practical reason of data availability. One thing needs to be noticed: since the increase of unemployment rate reflects the deterioration of one country's economy, a high unemployment rate therefore denotes negative macro-economic effects. Yet, an increase in unemployment rate might be considered as « good news » by some investors (Boyd et al., 2005).

## 4. Data and Preliminary Numerical Results

The data used in this study are obtained from DataStream from January 2001 to December 2014 (DataStream, 2019)[1]. According to the Morgan Stanley Capital International Emerging Market Index, 24 developing countries qualify as emerging markets. We first selected the top 7: China, Brazil, India, Russia, Mexico, Indonesia, and Turkey, whose nominal GDP was greater than six hundred million USD in 2014 (International Monetary Fund, 2015). We then had to exclude Russia due to data non-availability, reducing the data comparison to 6 countries.

It should be emphasized that the raw data, even obtained from validated sources sometimes contain anomalous values. Several industries, such as consumer services (*SVS*) and oil and gas (*OIL*), have maximum returns larger than 100 percent. In contrast, the minimum return of consumer services (*SVS*) is smaller than -100 percent. Those phenomena are uncommon for the stock returns. Of course, the returns of consumer goods (*GDS*), consumer services (*SVS*), health care (*HEA*), oil and gas (*OIL*), and telecommunications (TELE) industries can be significantly negative or positive. Nevertheless, most values do concentrate near zero. Accordingly, unlike the normal distribution, the histograms of those industry returns have relatively high peaks and thin tails. This suggests that data of FTSE global price indexes contains fault values. In order to reduce the influence of anomalous 'reported' returns, we first use a widely accepted fraud detection method in order to investigate the problematic aspects of the data set. We then adjusted the data set by deleting all the abnormally repetitious values pointed out when applying Benford's laws (Ausloos et al., 2015).

[1] **Data Availability statement:**
The data that support the findings of this study are available from Datastream. Restrictions apply to the availability of these data, which were used under license, at the University of Leicester Library, for this study. Data are available from the authors with the permission of University of Leicester and Datastream, as obtained from https://www2.le.ac.uk/library/find/databases/d/datastream

Table 2 reports the adjusted data set's descriptive statistics. Notice (1) almost all of the industries' mean returns are positive except for consumer services industry (*SVS*); this situation partly reflects that industries of the developing world generally enjoy a sustainable growth; (2) the average returns of several industries, e.g. the mean return of consumer services (*SVS*), are negative; (3) the average industry returns of basic materials (*MATS*), consumer goods industry (*GDS*), financials (*FIN*), health care (*HEA*), industrials (*INDU*) and industrials (*INDU*) are comparatively higher than the mean industry returns of consumer services (*SVS*), oil and gas (*OIL*), technology (*TECH*), and telecommunications (*TELE*). This is interpreted as follows: the former six industries are the booming industries of the developing world, while the latter four industries are still in an initial development stage; (4) the volatilities of industry returns seem to belong to a normal distribution; this is interpreted as resulting from the eliminations of the extreme values; all the values of standard deviations are close to 5, i.e. all the industries' returns fluctuate in « normal ranges »; (5) Table 2 also indicates that the statistical properties of GIS  and LIS are distinct from each other. This phenomenon implies our expectations, i.e. those two kinds of investor sentiments might affect the expected stock returns in different ways.

Table 3 reports measures of the correlations between the variables. We highlight the following aspects: (a) the correlations between industry returns are all positive but have different extents: some industries' returns (e.g. basic materials (*MATS*)) are closely related to other industries' returns, while several industries' returns (e.g. technology (*TECH*)) have weak correlations with other industry returns; (b) the relationships between industry returns and local sentiments (*GIS*) are all positive but weak; (c) the industry returns of consumer services (*SVS*), financials (*FIN*), health care (*HEA*), industrials (*INDUS*), and oil and gas (*OIL*) are positively linked with the global sentiments (*GIS*), while the other five industries' returns have negative correlations with the global sentiments (*GIS*); (d) both global sentiments (*GIS*) and local sentiments (*LIS*) have a positive but weak correlation; (e) both kinds of investor sentiments are positively related to business cycles (*BC*).

## 5. Empirical results

## 5.1 Local investor sentiments correlations to the stock returns

The regression analysis results for the ten industries are reported in Table 4. The returns of basic materials (*MATS*), consumer goods (*GDS*), financials (*FIN*) and industrials (*INDUS*) have a significant positive correlation with LIS, i.e. positive and negative emotions of local traders increase or decrease those industries' expected returns respectively. Furthermore, the coefficients of basic materials (*MATS*) and financials (*FIN*), which are comparative larger than other coefficients, reflect the fact that those two industries have relatively strong reactions to LIS. Besides, the industry returns of consumer services (*SVS*), health care (*HEA*), oil and gas (*OIL*), technology (*TECH*), telecommunications (*TELE*), and utilities (*UTIL*) seem to have no significant relationship with LIS.

Additionally, except for oil and gas (*OIL*), the other five industries' expected returns are insignificantly connected with both GIS and LIS, i.e. investor sentiments have no obvious influence on those expected industry returns. Oil and gas companies are exposed to systematic asset price risks, (Hamilton, 1983; Burbidge and Harrison, 1984; Sadorsky, 1999; Gupta and Banerjee, 2019) both oil prices and gasoline prices (Bacon, 1991; Deltas, 2008), and their investment behavior respond to the dynamic political and market environments (Mohn, K., & Misund, B., 2011).). Such results indicate that most global sentiments coefficients are positive, but *GIS* has an insignificant impact on industry returns. Notice that this is inconsistent with the findings of Chen et al. (2013). ~~Only~~ The effect of the investor sentiment is more significant on resource-based industries: basic materials (*MATS*) and industrials (*INDUS*) expected returns for those two industries' stocks could be significantly influenced by the fluctuations of global investors' emotion. These stocks are more resistant because the investors may raise their investment to avoid risks.

This finding implies that the world-wide optimism of stock investors could only slightly increase the expected industry returns, the negative impacts of GIS, i.e. pessimism, is also limited.

## 5.2 Local investor sentiments predict the stock returns but global sentiments do not

We find that local sentiments have heterogeneous effects on different industries' stock returns for emerging countries. The LIS significantly and positively affect the expected returns of only several specific industries. More precisely, for basic materials (*MATS*), consumer goods (*GDS*), financials (*FIN*) and industrials (*INDUS*), pessimistic LIS negatively influence the expected industry returns, while optimistic LIS produce positive impacts. Our research conclusions are consistent with Brown and Cliff (2005), Kumar and Lee (2006), Baker and Wurgler (2006, 2007) and Chen et al. (2013). In agreement with Baker and Wurgler (2006), in particular, optimistic and pessimistic LIS would lead to the overvaluation and undervaluation of stocks respectively, and therefore increase and decrease the expected stock returns, respectively. Besides, we find that for many industries, such as health care (*HEA*), consumer services (*SVS*), oil and gas (*OIL*), technology (*TECH*), telecommunications (*TELE*) and utilities (*UTIL*), the expected industry returns have no significant correlation with LIS.

The corresponding interpretations vary with the mentioned industries. For health care (*HEA*) and telecommunications (*TELE*), two industries closely related to the interests of countries, many dramatic changes in those industries would bring a loss of national benefits. Hence, one can imagine that the regulators in these emerging economies try to keep those industries in control and prevent the industry returns from influences of investor sentiments. Of course, stocks of oil and gas (*OIL*) industry are regarded as traditional bond-like equities which have stable returns; thus, their expected returns are less influenced by investor sentiments.

One should stress that, for emerging countries, LIS are more influential than GIS. The interpretation goes as follows: because in the emerging stock markets, local investors could trade shares more freely than foreign investors, the expected industry returns could only demonstrate fluctuations induced by LIS. Interestingly, technology (*TECH*) is negatively (but not significantly) related to local sentiments; Baker and Wurgler (2006, 2007) provided some possible explanations which we consider to be realistic, for the phenomenon. They stated that those speculative stocks are difficult to price and thus hard to arbitrage; therefore, higher optimism periods of the stock market are expected to produce lower future returns of the stocks.

However, *GIS* cannot predict most of the industries' expected returns, while numerous existing studies, e.g. Chen et al. (2013), suggest that global sentiments could significantly affect the expected return of several specific industries. This discrepancy between findings might be generated by the differences between research objects. In our research, we take emerging markets as the investigated object, rather than developed markets. Those emerging stock markets completely differ from the well-developed markets. On one hand, the governments play significant roles in the stock markets of emerging countries; thus the stock markets' degrees of freedom are decreased (Harvey, 1995; Claessen, Dasgupta, and Glen, 1995). Take the Chinese stock market for example: the Chinese government has issued many regulations to limit capital investment from foreign country 'for security reasons'. Additionally, external investors are restricted to buy Chinese equities; therefore, changes of GIS could not bring tremendous impacts on emerging countries' expected stock returns. On the other hand, due to the limitations of investment knowledge, local stock traders might not respond notably or quickly to some foreign financial information; they are more sensitive to domestic news. Hence, the expected returns would not reflect the changes in the GIS. The foreign investor, however, may face formal or informal barriers when trading (Serra, 2000) because they might be more informative and efficient (Vo, 2017, Syamala and Wadhwa, 2019) but exposed to more risk of stock price crash (Vo, 2018). The different types of investors, i.e. individual and institutional investors may also exert dynamic behavior due to the different momentums (Koesrindartoto et al, 2020). Due to such reasons, global sentiments could not be a significant explanatory variable for expected industry returns. This situation is thus inconsistent with the cases of developed stock markets, leading to the observed difference in behavior.

Notice that our empirical tests suggest that linear regression models of *LIS*, *GIS*, and *BC* factors might not significantly explain the fluctuations of expected industry stock returns. A possible explanation is that global sentiments are not a significant explanatory variable while local sentiments are not adequate to demonstrate all the changes of industry future returns. We suggest that a more complete anticipation of stock returns, for these markets, it seems to us, needs to bring other factors, such as the market status, company financial statements and risk-free rates, or even IPO news (Kim and Byun, 2010), into consideration. Moreover, other investor sentiments indexes might be invented (Bu and Li, 2014) and thereafter tested. It would be also interesting to examine another type of causality, whether the stock market fluctuations, in such (emerging) countries and for this set of different industry sectors, influence the investor sentiments (Chen et al., 2003) rather than the contrary, as assumed here. Most likely the time correlations will have different spans; the investor, local or global, sentiment reaction might also have to be measured through other means, but the matter is another interesting open door.

## 6. Conclusions

A survey of the literature implies that investor sentiment's influence on expected stock returns is an open field for research. Plenty of previous empirical analyses suggest that the fluctuations of investor sentiments can significantly sway future stock returns. However, we point out that only few studies investigate investor sentiment's heterogeneous effects on different industrial sectors' stock returns. Moreover, most of the existing research has focused on developed countries' stock markets as research object, leaving aside investigations of stock markets of emerging countries. Thus, in our study, we do analyze investor sentiment's (both local sentiments and global sentiments) heterogeneous effects on different industries' stock returns. To this aim, we follow Schmeling (2009) and Chen et al. (2013) research approaches and build a fixed effect linear regression model in order to study the correlations between investor sentiments and different industries' expected returns. Consequently, we selected six most significant emerging countries to represent the whole 'developing world'. Accordingly, the State Street Investor Confidence Index, each country (« local ») consumer confidence index, each country unemployment rate and the FTSE global price indexes are employed as the proxies for global sentiments, local sentiments, business cycle and expected industry returns respectively. After a group of statistical tests and necessary data adjustments, a panel data set has been constructed to perform an empirical analysis.

Sometimes confirming previously published conclusions about LIS's impacts, our empirical results also indicate some unique features of investor sentiments influencing mechanism for developing countries' stock markets. Unlike mature stock markets, the expected returns of emerging markets' stocks are insignificantly related to global sentiments; in other words, changes of global investors' emotions do not bring apparent changes of future stock returns. We conclude that likely due to government interventions and restrictions on foreign capital investment, emerging markets are insensitive to the impacts of global sentiment. For local sentiment-expected industry returns nexus, we find that local sentiments only positively and significantly affect the expected returns of basic materials (*MATS*), consumer goods (*GDS*), financials (*FIN*) and industrials (*INDUS*). Nevertheless, the other six industries' expected returns have no significant correlation with LIS. This finding, which is consistent with the general consensus in the former literature, confirms that investor sentiments have heterogeneous influences on different industries' stock returns. Moreover, the results show that it is inadequate to only use local sentiments, global sentiments and business cycle factor to anticipate future stock returns. We hold the opinion that other necessary factors, such as the market status, company financial statement and risk-free rate should be taken into account to conduct an effective anticipation of stock returns.

We confirm that this article contains a Data Availability Statement (see section 4 )
We confirm that we include how available data can be (and was) obtained ; see references section :
DataStream (2019)

**Table 1.** About previous studies on investor sentiments and stock returns

This table proposes some information about several previous studies (in chronological order) on the relationship between investor sentiments and stock returns. The selection is made in order to provide perspectives; other papers are discussed in the text.

| Publish Year | Author(s) | Research Object(s) | Method | Data | Research Objective | Research results |
|---|---|---|---|---|---|---|
| 2002 | Simpson and Ramchander | Australia | Regression | Monthly data 1985-1999 | Investigating how investor sentiments influences the time varying part of discounts and premia of the First Australian Fund. | Investor sentiment has huge influence on the change of premia on the First Australia fund. |
| 2004 | Chelley-Steeley and Siganos | UK | OLS | Monthly data 1975-2001 | Whether macroeconomic variables or market sentiment is more influential to momentum profits. | Macroeconomic variables could affect momentum profits, while market sentiment could also influence the profitability of momentum trading strategy. |
| 2005 | Brown and Cliff | U.S. | Regression | Monthly data 1963-2000 | Investigating that whether a survey of investor sentiment have the ability to estimate market return over following years. | Future return has a negative relationship with investor sentiment. |
| 2006 | Baker and Wurgler | U.S. | Regression | Monthly data 1962-2001 | Whether sentiment has the cross-sectional effects on stock returns or not. | Investor sentiment is more likely to influence smaller, young, less stable, less profitable and non-dividend-paying stocks. |
| 2006 | Tsuji | Japan | Regression | Monthly data 1994-2001 | Investigates investor sentiment's predictability for expected stock return and its properties in Japan. | Investor sentiment has predict power for short-term expected stock return, while there are no "naïve extrapolation" and "salience effect" in Japan. |
| 2009 | Canbas and Kandir | Turkey | VAR model | Monthly data 1997-2005 | Analyzing the correlation between investor sentiment and stock return. | Investor sentiment might not have the ability to anticipate stock return. |
| 2009 | Schmeling | Developed countries. | Regression | Monthly data 1985-2004 | Using the international evidence to prove the connection between investor sentiment and stock return. | Investor sentiment could predict the stock return and those countries whose culture encourages investment behavior are more vulnerable to the influence of sentiment. |
| 2010 | Kaplanski and Levy | Aviation disasters | Regression | Daily data 1950-2007 | Aviation's influence on stock price. | Less stable firms are more vulnerable to investor sentiment. Investor sentiment is a significant factor in estimating future returns. |
| 2010 | Kurov | U.S. | OLS | Monthly data 1990-2004 | Investigating monetary policy's impacts on investor sentiment. | In bear markets, investor sentiment and monetary policy decision have a strong correlation. |
| 2011 | Yu and Yuan | U.S. | GARCH (1,1) | Monthly data 1963-2004 | The correlation between investor sentiment and the mean-variance tradeoff. | In the low-sentiment periods, market's expected extra return positively link with its conditional variance, while in the high-sentiment periods, there is no connection between them. |
| 2012 | Stambaugh, Yu, and Yuan | Portfolios | Regression | 11 anomalies over Aug. 1965 to Jan. 2008 | Effect of short-sale impediments. | Sentiment has no relation to returns on the long legs strategies, but short leg strategies are profitable following high sentiment |
| 2013 | Chen, Chen, and Lee | 11 selected Asian countries | Regression and panel threshold model | Monthly data 1996-2010 | Analyzing the global and local investor sentiment's correlation with industry returns with both linear and non-linear methods. | Both global and local sentiment have asymmetric influences on industry returns and relationship between expected industry return and investor sentiment varies with different sentimental intervals. |
| 2014 | Huang, Yang, Yang and Sheng | China | Regression | Monthly data 2005-2013 | Investigating the influence of investor sentiment on a certain industry's stock return. | In the current period, investor sentiment has a positive relationship with return of an industry, while they are negatively correlated in one lag period. |
| 2014 | Oprea and Brad | Romania | Regression | Monthly data 2002-2011 | Investigating the influence of investor sentiment on stock prices. | Investor sentiment has a positive relationship with returns from small stocks but not from large stocks. |

**Table 2.** Descriptive statistics of the adjusted data set used in this study

This table reports the (rouned) main descriptive statistics of variables after the deletion of abnormal repetitious numbers. The variables are ten industries' monthly stock return, global investor sentiment, local investor sentiment of six countries and business cycle factor.

| Variable | Obs | Mean | Std. Dev. | Min | Max |
|---|---|---|---|---|---|
| MATS | 1080 | 0.2883 | 5.9418 | -35.9707 | 23.0324 |
| GDS | 906 | 0.4373 | 4.8700 | -21.6857 | 24.1562 |
| SVS | 992 | -0.0240 | 5.5279 | -62.0864 | 19.4596 |
| FIN | 1075 | 0.3989 | 5.1834 | -26.3445 | 22.4560 |
| HEA | 724 | 0.2664 | 4.3248 | -17.6841 | 18.9232 |
| INDUS | 1007 | 0.2619 | 5.3034 | -34.2964 | 21.0071 |
| OIL | 816 | 0.0515 | 5.3762 | -28.3126 | 20.9206 |
| TECH | 449 | 0.0241 | 5.9233 | -25.3684 | 45.1308 |
| TELE | 1015 | 0.0987 | 4.8096 | -35.1788 | 22.9844 |
| UTIL | 745 | 0.3756 | 4.8605 | -19.1293 | 26.0514 |
| GIS | 1080 | 109.346 | 14.120 | 80.50 | 141.80 |
| LIS | 1075 | 106.443 | 19.028 | 9.60 | 170.18 |
| BC | 980 | 6.8956 | 2.8558 | 2.15 | 13.900 |

**Table 3.** Correlation analysis between variables in this study

The table shows the Pearson correlation coefficient value between ten different industry sectors' stock returns, global investor sentiments, local investor sentiments and business cycle in the 2000 to 2014 period. All data involved is collected from DataStream

| | MATS | GDS | SVS | FIN | HEA | INDUS | OIL | TECH | TELE | UTIL | GIS | LIS | BC |
|---|---|---|---|---|---|---|---|---|---|---|---|---|---|
| MATS | 1 | | | | | | | | | | | | |
| GDS | 0.7091 | 1 | | | | | | | | | | | |
| SVS | 0.7484 | 0.7477 | 1 | | | | | | | | | | |
| FIN | 0.8208 | 0.6874 | 0.7140 | 1 | | | | | | | | | |
| HEA | 0.6120 | 0.6318 | 0.5907 | 0.5498 | 1 | | | | | | | | |
| INDUS | 0.8686 | 0.7576 | 0.7552 | 0.8583 | 0.6161 | 1 | | | | | | | |
| OIL | 0.8157 | 0.6121 | 0.6647 | 0.7654 | 0.5647 | 0.8112 | 1 | | | | | | |
| TECH | 0.5521 | 0.3644 | 0.3602 | 0.4731 | 0.3816 | 0.5363 | 0.4419 | 1 | | | | | |
| TELE | 0.7011 | 0.5459 | 0.6217 | 0.6890 | 0.4859 | 0.7150 | 0.7054 | 0.5158 | 1 | | | | |
| UTIL | 0.7593 | 0.6043 | 0.6463 | 0.7710 | 0.5321 | 0.7772 | 0.7345 | 0.5288 | 0.6646 | 1 | | | |
| GIS | 0.1881 | 0.0640 | 0.1221 | 0.1170 | 0.0593 | 0.0975 | 0.0739 | 0.1219 | 0.1166 | 0.1431 | 1 | | |
| LIS | -0.0220 | -0.0337 | 0.0619 | 0.0107 | 0.0121 | 0.0141 | 0.0082 | -0.0536 | -0.0729 | -0.0451 | 0.0028 | 1 | |
| BC | 0.0679 | -0.0320 | 0.0278 | 0.0038 | 0.0052 | 0.0076 | 0.1163 | -0.0548 | 0.0362 | 0.0032 | 0.2420 | 0.1986 | 1 |

### Table 4. Regression coefficients results of fixed effect models

This table reports the regression coefficient values of the fixed effect models for each explanatory variable (see text). The dependent variables are the expected returns for ten different industry sectors stocks. Numbers in parentheses are the standard errors on the coefficients for the correspondent explanatory variables. The *, **, and *** indicate that the correspondent variable has passed the 95%, 99% and 99.9% t-statistic test respectively.

| VARIABLES | MATS | GDS | SVS | FIN | HEA |
|---|---|---|---|---|---|
| GIS | 0.0579*** | 0.0125 | 0.0201 | 0.0244 | -0.0089 |
| | (0.015) | (0.015) | (0.022) | (0.013) | (0.011) |
| LIS | 0.0564*** | 0.0350** | 0.0334 | 0.0469*** | 0.0135 |
| | (0.013) | (0.012) | (0.019) | (0.011) | (0.009) |
| BC | 0.4043*** | 0.2016 | 0.2392 | 0.2364* | 0.0774 |
| | (0.112) | (0.104) | (0.162) | (0.097) | (0.085) |
| « Constant » | -14.7369*** | -6.0111** | -7.0518 | -8.8560*** | -0.7183 |
| | (2.528) | (2.290) | (3.654) | (2.187) | (1.793) |
| R-square | 0.041 | 0.013 | 0.005 | 0.022 | 0.004 |

| VARIABLES | INDUS | OIL | TECH | TELE | UTIL |
|---|---|---|---|---|---|
| GIS | 0.0264* | 0.0026 | 0.0197 | 0.0207 | 0.0186 |
| | (0.013) | (0.017) | (0.018) | (0.012) | (0.013) |
| LIS | 0.0250* | 0.0299 | -0.0053 | 0.0188 | 0.0147 |
| | (0.012) | (0.015) | (0.026) | (0.010) | (0.010) |
| BC | 0.1997* | 0.4211*** | -0.1659 | 0.2009* | 0.1550 |
| | (0.099) | (0.122) | (0.140) | (0.088) | (0.091) |
| « Constant » | -6.5200** | -6.1659* | -0.1938 | -5.3780** | -4.3823* |
| | (2.224) | (2.831) | (3.258) | (2.002) | (2.047) |
| R-square | 0.012 | 0.017 | 0.004 | 0.011 | 0.01 |

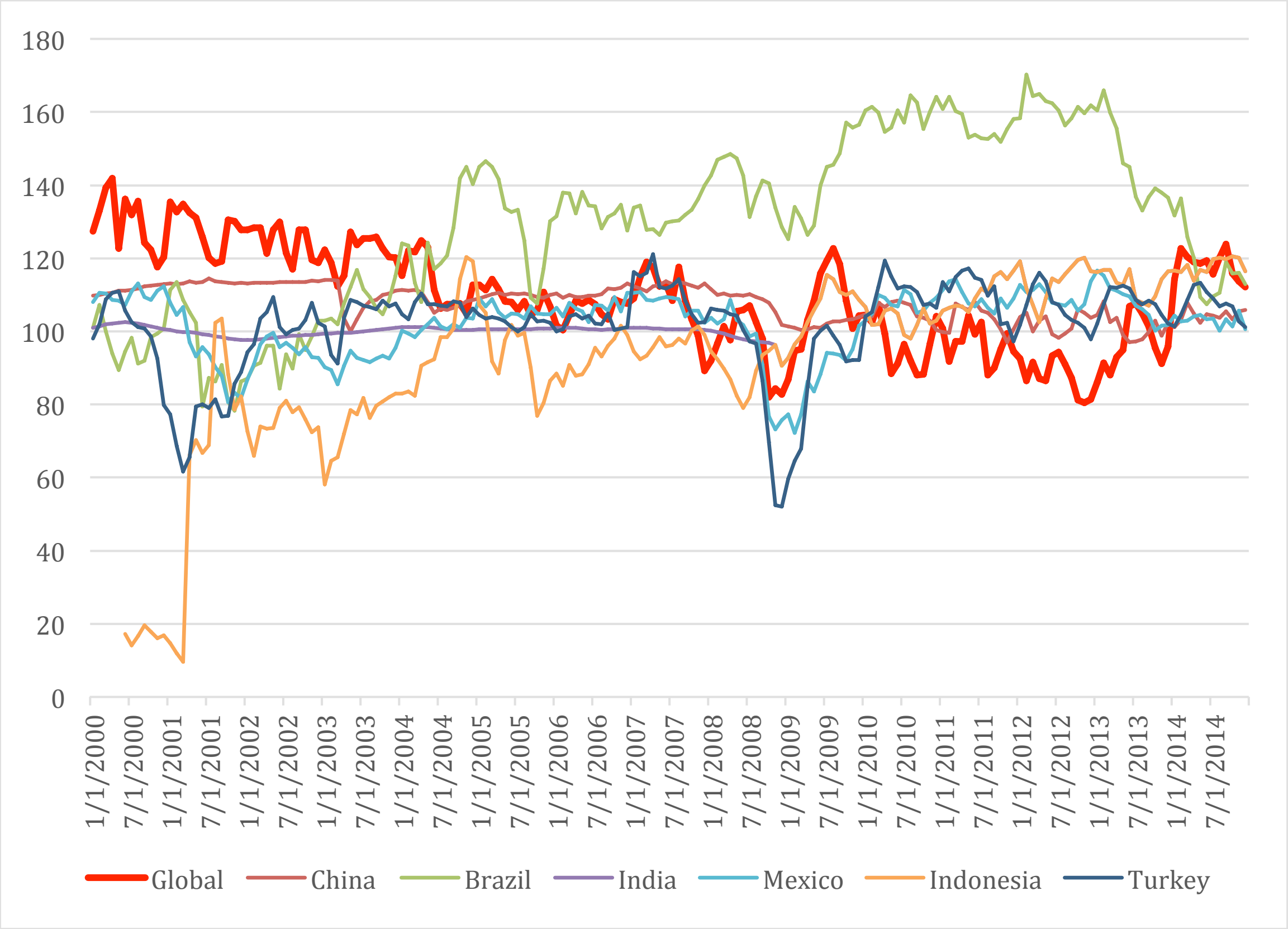

**Figure 1.** Global and local investor sentiment

The line graph illustrates changes in the global and in six selected countries' investor sentiments indices. The indicators are each country's official consumer confidence index and the State Street Investor Sentiment Index (https://www.investopedia.com/terms/s/state-street-confidence-index.asp). The area between the two black lines represent the financial crisis time between September 2007 and June 2009.